# Cavity-enhanced circular dichroism in a van der Waals antiferromagnet

Shu-Liang Ren[†], Simin Pang[†], Shan Guan, Yu-Jia Sun, Tian-Yu Zhang, Nai Jiang, Jiaqi Guo, Hou-Zhi Zheng, Jun-Wei Luo, Ping-Heng Tan, Chao Shen*, and Jun Zhang*

ABSTRACT: Broken symmetry plays a pivotal role in determining the macroscopic electrical, optical, magnetic, and topological properties of materials. Circular dichroism (CD) has been widely employed to probe broken symmetry in various systems, from small molecules to bulk crystals, but designing CD responses on demand remains a challenge, especially for antiferromagnetic materials. Here, we develop a cavity-enhanced CD technique to sensitively probe the magnetic order and broken symmetry in the van der Waals antiferromagnet $FePS_3$. By introducing interfacial inversion asymmetry in cavity-coupled $FePS_3$ crystals, we demonstrate that the induced CD is strongly coupled with the zig-zag antiferromagnetic order of $FePS_3$ and can be tuned both spectrally and in magnitude by varying the cavity length and $FePS_3$ thickness. Our findings open new avenues for using cavity-modulated CD as a sensitive diagnostic probe to detect weak broken symmetries, particularly at hidden interfaces, and in systems exhibiting hidden spin polarization or strong correlations.

The investigation of magnetic order and broken symmetry in two-dimensional (2D) magnetic materials has attracted widespread attention[1-3] due to its importance for the development of 2D magnetic-related devices, such as magnetic storage and on-chip quantum communication.[4-7] Therefore, it is imperative to directly probe and manipulate the magnetic order and broken symmetry in 2D magnets. Techniques like natural circular dichroism (NCD) and magnetic circular dichroism (MCD) spectroscopy, which measure the difference in reflected or transmitted light intensity between left and right circularly polarized (LCP and RCP) light,[8-11] are sensitive probes for broken symmetry. NCD requires broken inversion symmetry, while MCD emerges in magnetic materials with broken time-reversal symmetry, typically induced by an external magnetic field oriented along the direction of light propagation.[12] Notably, the behavior of NCD and MCD is different for light propagation reversal: NCD accumulates in a cavity, enhancing the signal, while the MCD accumulated in the forward direction is compensated by backward propagation, making them less sensitive in cavities. The cavity-enhanced effect for NCD and MCD is different from the case for optical rotation and Faraday rotation,[13] as schematically shown in Figure S1.

In 2D materials, typical magnetic orders include ferromagnetism and antiferromagnetism.[14] Compared to ferromagnetic materials, 2D antiferromagnetic (AFM) materials, such as $FePS_3$, offer higher response frequency and no stray fields, making them promising for next generation of AFM spintronics devices.[15-17] $FePS_3$, an Ising type van der Waals AFM, can be easily exfoliated to atomic thickness, retaining long-range magnetic order even at the monolayer limit, providing a unique platform for conducting fundamental studies related to 2D AFM properties.[18, 19] Previous studies have demonstrated several attractive phenomena in $FePS_3$, including a giant out-of-plane magnetic anisotropy,[20] near-unity linear dichroism (LD),[21] giant surface second-harmonic generation (SHG) coupled to nematic orders,[22] divergence behavior of the demagnetization time near the magnetic phase transition investigated by magnetic linear birefringence (MLB) and optical pump-probe spectroscopy,[23] and magnon-phonon coupling studied by Raman spectroscopy.[24] Bulk FePS$_3$ has a monoclinic structure (space group *C2/m*), which is inversion symmetric, with a band gap of approximately 1.5 eV.[25] As shown in Figure 1a, iron (Fe) atoms are octahedrally coordinated by six sulfur (S) atoms at the vertices and form a honeycomb lattice in the *ab* plane. A pair of phosphorus (P) atoms is located at the center of each hexagon. The magnetic properties of FePS$_3$ are determined by the Fe *d*-electron. Below $T_N$, FePS$_3$ exhibits zig-zag AFM order.[18] In the hexagonal honeycomb plane composed of Fe atoms, nearby zig-zag chains have opposite spin orientations, i.e., spin up and spin down perpendicular to the honeycomb plane, respectively, as shown by the red and blue arrows in Figure 1b. The adjacent planes are coupled antiferromagnetically. In the paramagnetic (PM) state ($T>T_N$), FePS$_3$ is structurally centrosymmetric. Below $T_N$, FePS$_3$ undergoes a magnetic phase transition to the AFM state, accompanied by a simultaneous structural phase transition. Across the phase transition, the crystal structure remains monoclinic, but the monoclinic angle experiences a subtle and abrupt change.[26,27] Figure 1b shows that the zig-zag AFM order preserves the inversion symmetry. Therefore, for FePS$_3$ in both the PM and AFM states, the NCD vanishes. Additionally, despite the broken time-reversal symmetry due to the zig-zag AFM,[28] the combined half-unit cell translation and time-reversal symmetry preserves the band Kramers degeneracy, resulting in zero MCD. This absence of NCD and MCD in fully compensated collinear AFM FePS$_3$ prevents them from being appropriate probes. Inducing a nonzero CD in the FePS$_3$ system and enhancing it is essential for detecting antiferromagnetism and advancing applications based on antiferromagnets.

In this report, we studied the magnetic order and broken symmetry of FePS$_3$ using cavity-enhanced CD spectroscopy. By constructing an asymmetric interface between FePS$_3$ and the substrate, we induced a small, nonzero NCD, which originates from an interfacial spin-orbit magnetic field ($B_0$) and is strongly coupled to the zig-zag AFM order of FePS$_3$. This induced NCD is highly sensitive to both the structural and spin symmetry of FePS$_3$ and can be modulated by adjusting the cavity length or FePS$_3$ thickness. Notably, below the Néel temperature ($T_N$), where spin and interfacial asymmetries more strongly break inversion symmetry, the CD signal is significantly enhanced, providing a clear indicator of magnetic phase transitions and spin ordering. These findings present cavity-enhanced CD as a promising tool to explore antiferromagnetism and broken symmetries in materials where conventional probes struggle due to the absence of net magnetization.

When the material has low light absorption, interference effect becomes prominent at a certain thickness. Figure 1c shows a schematic of multiple reflection and optical interference in a multilayer structure consisting of air, FePS$_3$ flake, SiO$_2$, and Si substrate, forming an optical cavity. This cavity enhances the CD by attenuating the reflected light due to the interference effect.[29] Since the samples were exfoliated on opaque substrates, we used a reflection setup rather than transmission. Reflectance CD measurements were conducted using a polarization modulation technique in our home-built microscopic system, as schematically shown in Figure 1d (detailed in Section 3 of the Supporting Information). Figure 1e shows the optical constants $n$ and $k$ of the bulk FePS$_3$ measured with an ellipsometer at room temperature. Mathematically, CD is defined as

$$CD = \frac{R_+(E) - R_-(E)}{[R_+(E) + R_-(E)]/2} \propto \frac{1}{R}\frac{dR(E)}{dE}\Delta n, \tag{1}$$

where $R_+$ ($R_-$) is the reflectance of LCP (RCP) light, $R = (R_+ + R_-)/2$ is the average reflectance without polarization, and $\Delta n$ is the refractive index difference between LCP and RCP lights.[30-32] More details about the derivation of eq 1 can be found in Section 4 of the Supporting Information.

The CD spectrum can be approximately as the first-order differential of the modulated reflectance spectrum. An obvious CD appears at the minimum of the reflectance spectrum, which is proportional to $\Delta n$. The cavity-enhanced effect enables the detection of small $\Delta n$.

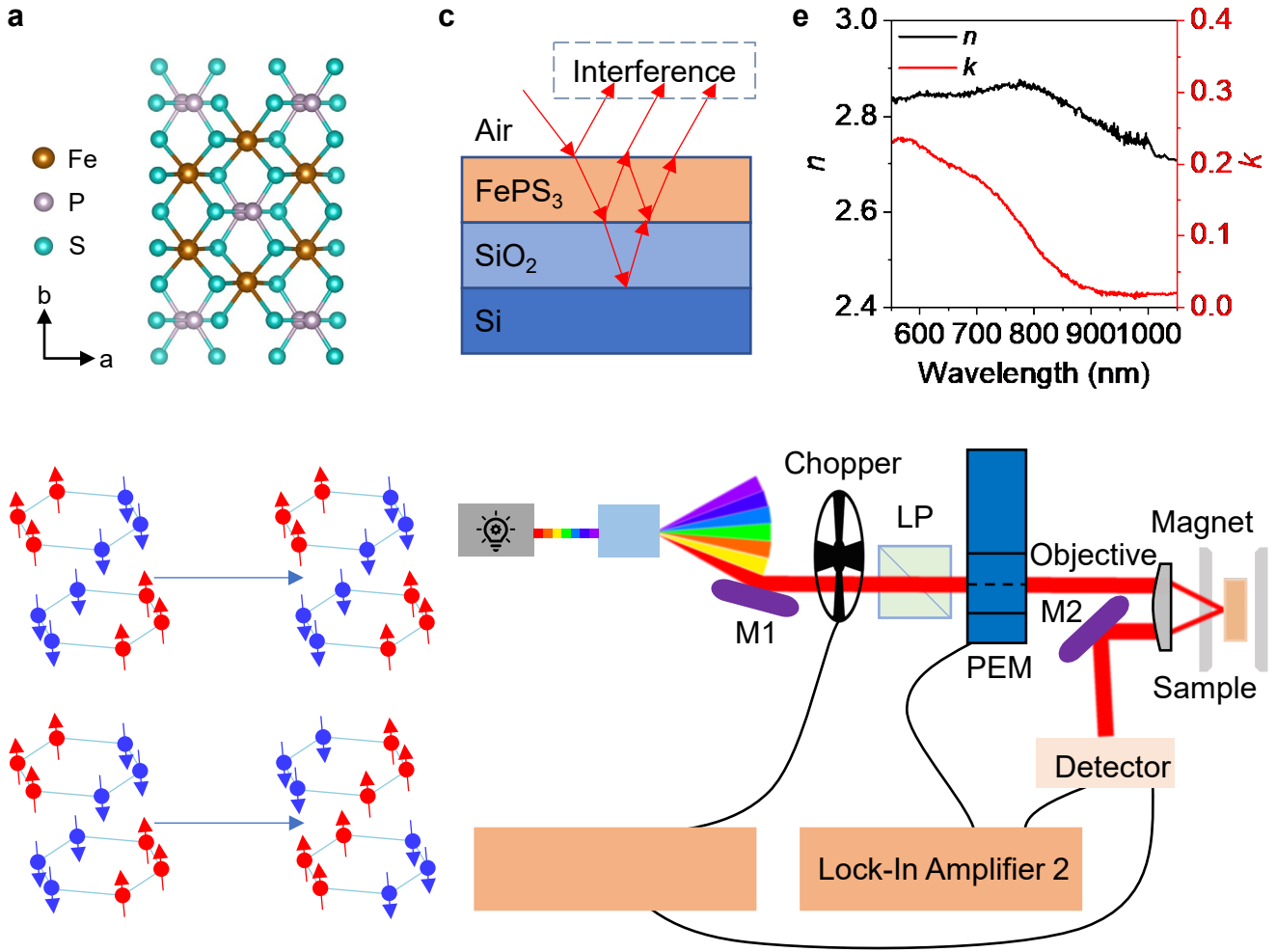

**Figure 1.** The structure, magnetic, and optical properties of FePS$_3$, together with the schematic optical interference and experimental setup for CD measurements. (a) Top view of the atomic lattice of the FePS$_3$ crystal. Brown, purple, and mint green spheres represent Fe, P, and S atoms, respectively. (b) Symmetry of the spin structure on the honeycomb lattice of Fe$^{2+}$ at 0 T. Red and blue arrows represent spin up and spin down, respectively. The zig-zag AFM order ($T<T_N$) in FePS$_3$ preserves the inversion symmetry (leading to a vanishing NCD) while breaking the time-reversal symmetry. However, the combination of the half-unit cell translational operation with the time-reversal operation is a good symmetry, resulting in a zero MCD. (c) Schematic of multiple reflection and optical interference in the cavity-enhanced CD spectra measurements. (d) Schematic of the experimental setup for CD measurements. The M, LP, and PEM represent mirror, linear polarizer, and photoelastic modulator, respectively. (e) Optical constants of bulk FePS$_3$ measured with an ellipsometer at room temperature.

FePS$_3$ with a specific thickness ($d$) can form an internal cavity, enhancing light-matter interaction and, thus, the CD. The cavity modes occur at wavelengths ($\lambda$) that satisfy $d = p \frac{\lambda}{2n_{eff}}$, where $p$ represents the cavity mode order and $n_{eff}$ denotes the effective refractive index of the stacked FePS$_3$/SiO$_2$/Si.[21] To demonstrate the cavity-enhanced CD effect, reflectance (red) and CD (blue) spectra were measured on FePS$_3$ with six distinct thicknesses ($d$) exfoliated on 90 nm SiO$_2$/Si substrate under -3 T and 300 K (symbols in Figure S2). As the thickness increases, interference effects become more pronounced, with more characteristic signals in the reflectance spectra. The minimum reflectance corresponds to different orders of the cavity mode resonances, marked by their respective order numbers ($p$). The CD extremum, appearing on both sides of the corresponding minimum reflectance, also intensifies with increasing thickness and shifts towards longer wavelengths, indicating the interference effect's role in enhancing CD. The simulated spectra were denoted by the solid lines in Figure S2. To match the experimental results, the Δ$n$ in eq 1 was estimated as about 0.002 for all wavelengths in our calculations, though it should be wavelength-dependent. As shown in Figure S2, the experimental reflectance and CD spectra (symbols) mostly agree well with the simulated results (solid lines), particularly for thicker samples. However, a minor discrepancy exists between the experimental and simulated CD spectra for wavelengths below 827 nm (above the bandgap of 1.5 eV). In the CD spectrum calculations, we assumed that the CD primarily arises from the different responses of the real part of the complex refractive index ($n$) to the two circularly polarized lights, ignoring the effect of the weak imaginary part ($k$), which is valid when $n \gg k$, i.e., for wavelengths above 827 nm. Therefore, $k$ and $\Delta k$ cause deviations of the simulated results from the experimental spectra. Although the absolute values of the simulated reflectance and CD are less meaningful, our simulated results can still reproduce the shape characteristics of the experimental reflectance and CD spectra, and the intensity discrepancy would not affect our main conclusion.

To better understand the cavity-enhanced CD of the stacked $FePS_3/SiO_2/Si$, we calculated the reflectance and CD spectra for $FePS_3$ thicknesses from 50 nm to 1800 nm at 10 nm intervals. The calculated reflectance and CD intensity contour plots are shown in Figure 2a, c, respectively. The linearly dispersive branches of the resonances from bottom to top correspond to the 1st to 12th order cavity modes. The experimentally measured resonance positions of reflectance and CD extracted from Figure S2 are indicated by red symbols. For cavity modes of the same order, the CD signals are enhanced as the $FePS_3$ thickness increases due to more pronounced interference effects. Figure 2b, d display the reflectance and CD spectra for several typical thicknesses extracted from Figure 2a, c, respectively, intuitively displaying multiple orders of resonances. Notably, as shown in Figure 2c, $FePS_3$, with a thickness of less than 100 nm, exhibits no obvious CD as it is too thin to induce an interference effect and form an internal cavity for the photons within the detection wavelength region. Considering thin samples, we proposed that constructing an external cavity would also enhance the CD. We exfoliated an 80 nm $FePS_3$ on a distributed Bragg reflector (DBR) with a thickness of about 3 μm (Figure S3a). The CD is significantly enhanced here (Figure S3c), compared to the case on the 90 nm $SiO_2/Si$ substrate, where neither the internal nor external cavities are formed. Similarly, increasing the SiO$_2$/Si substrate thickness would also support an external cavity. Our results demonstrate that forming the internal or external cavity is an efficient method to enhance the CD in both thick and thin samples. The resonance magnitude and spectral position of the CD can be tuned simply by altering the thickness (cavity length) of the FePS$_3$ or substrate.

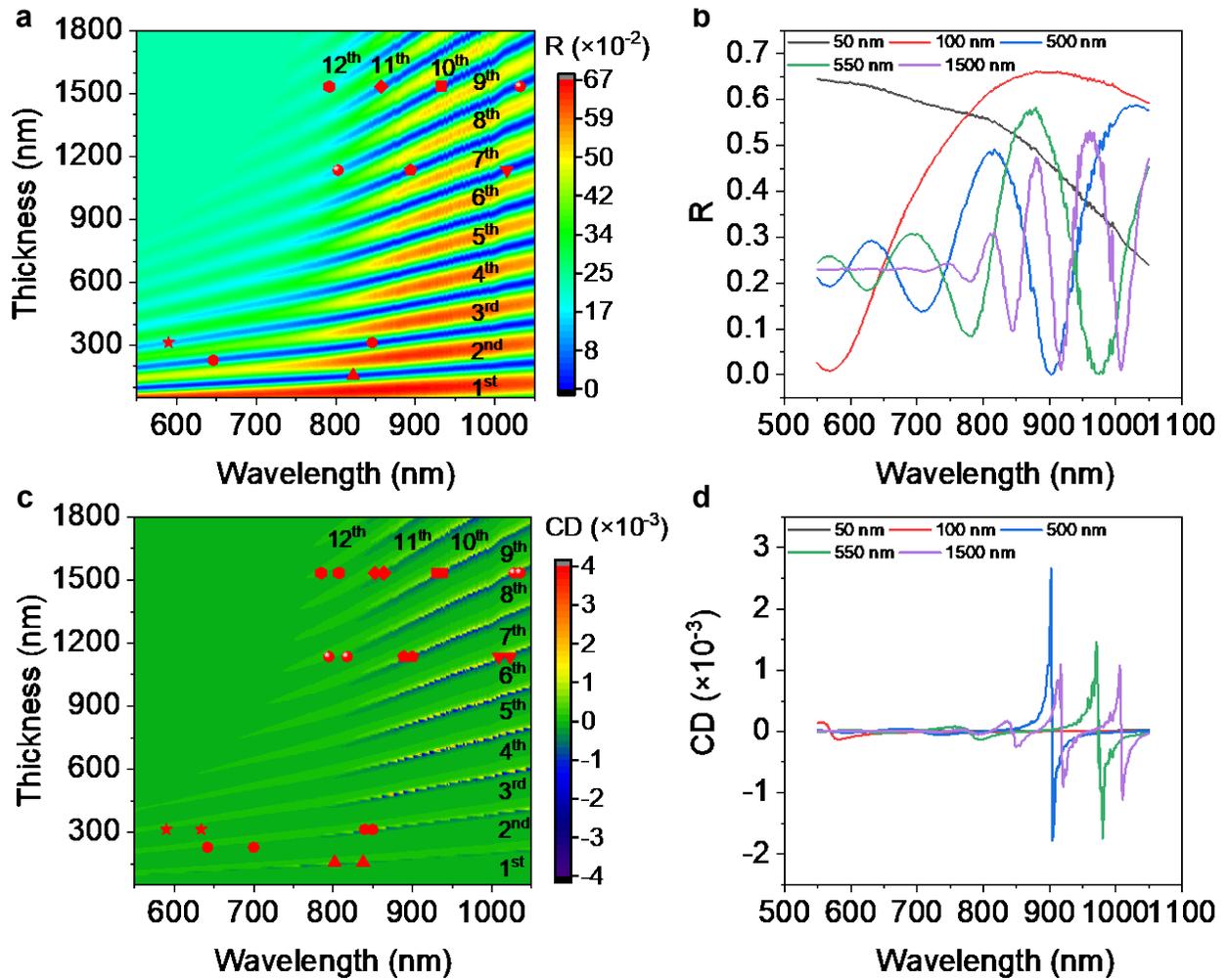

**Figure 2.** Simulated reflectance and CD spectra of FePS$_3$ on 90 nm SiO$_2$/Si substrate. (a) Reflectance and (c) corresponding CD intensity contour plot as a function of FePS$_3$ thickness, from 50 nm to 1800 nm at 10 nm intervals. The CD spectra were calculated from the reflectance spectra

according to Eq. (1) with $\Delta n \approx 0.002$. The red symbols in (a) and (c) represent the experimentally measured minimum reflectance and CD resonances with different orders extracted from Figure S2. (b) Reflectance and (d) CD spectra with several typical thicknesses (50 nm, 100 nm, 500 nm, 550 nm, and 1500 nm) extracted from (a) and (c), respectively.

At room temperature, FePS$_3$ is in the PM state, maintaining both inversion and time-reversal symmetry. Applying an external magnetic field would break the time-reversal symmetry, causing a nonzero MCD. The degree of broken time-reversal symmetry is proportional to the strength of the MCD. Since the thick sample possesses a more significant cavity-enhanced effect and thus a more apparent MCD, we investigated the magnetic field-dependent CD spectra at 300 K using the 1500 nm thick FePS$_3$ (S1 sample). The CD spectra show five interference characteristic wavelengths (around 1027 nm, 930 nm, 852 nm, 784 nm, and 721 nm), corresponding to the 8$^{th}$ to 12$^{th}$ order cavity modes (Figure 3a). Specifically, by constructing an asymmetric interface (FePS$_3$/SiO$_2$/Si substrate) that breaks the interface inversion symmetry, we successfully generate and detect the NCD at 0 T (green curve in Figure 3a), which should vanish in centrosymmetric PM FePS$_3$. Firstly, we can rule out that the finite CD at 0 T is attributed to the intrinsic ferromagnetism of FePS$_3$. As shown in Figure S4, the CD of the S1 sample measured from +0 T (down from the positive field) and -0 T (up from the negative field) was almost the same, indicating no hysteresis and FePS$_3$ is PM with no net magnetization at room temperature. We considered that due to the broken inversion symmetry, the interfacial spin-orbit magnetic field[33, 34] ($B_0$) generated at the FePS$_3$/SiO$_2$ interface is the predominant effect to induce the finite CD. Besides the FePS$_3$/90 nm SiO$_2$/Si, the DBR substrate formed by periodic arrangements of SiO$_2$ and Ta$_2$O$_5$ also exhibits weak CD at 0 T, owing to its interface inversion asymmetry (Figure S3b).

Under a certain magnetic field, the electron energy levels of different spins in the sample split, which can be detected using the MCD. The energy splitting ($\Delta E$) can be deduced from the MCD peak at energy $E_1$ and valley at $E_2$, that is $\Delta E = -\sqrt{e}w \frac{\Delta R_{MCD-peak}-\Delta R_{MCD-valley}}{2R_0}$, where $2w=|E_1-E_2|$ and $R_0$ is the reflectance at energy $E_0=(E_1+E_2)/2$.[30] As shown in Figure 3b, the $\Delta E$ of five resonant cavity modes extracted from Figure 3a depends linearly on the external magnetic field $B$, which can be well fitted by $\Delta E = A \times (B + B_0)$, where $A$ is the slope of the linear fit and $B_0$ represents the nonzero interfacial spin-orbit magnetic field. The $A$ and $B_0$ for three samples of different thickness, i.e., S1 (1500 nm), S2 (286 nm), and S3 (348 nm), are summarized in Figure 3c, d, respectively, with the full CD spectra of S2 and S3 shown in Figure S5. Clearly, both $A$ and $B_0$ depend on the wavelength and the thickness of sample. It has been reported that the frequency-dependent refractive index corresponding to RCP and LCP lights can be expressed as $n^{\pm}(\omega) \approx n(\omega) \pm \frac{dn}{d\omega}\frac{eH}{2mc}$, where $\omega$ ($c$) is the angular frequency (velocity) of light, $m$ is the electron mass, and $n(\omega)$ is the refractive index of the material in the absence of the magnetic field $H$.[35] Thus, the $\Delta E$, which is proportional to the refractive index difference ($\Delta n$), can be rewritten as $\Delta E \propto \frac{dn}{d\omega}\frac{e}{mc}H$. Therefore, the slope ($A$) of the energy splitting-magnetic field curve is proportional to $\frac{dn}{d\omega}$. According to the measured dispersion curve of the refractive index ($n$) for FePS$_3$ (Figure 1d), $n$ varies slightly at short wavelengths, which explains the small values of $A$ in the short wavelength regime (Figure 3c). Under the same nominal external magnetic field, the effective magnetic field ($B_{eff}$) experienced by a thin sample is greater than that by a thick sample, causing a more significant $\Delta E$ in the thin sample. Hence, the thin samples (S2 and S3) have relatively larger $A$ values compared to the thickest sample S1, as shown in Figure 3c. However, note that the $A$ we gave here is the nominal slope due to the difference in the effective magnetic field ($B_{eff}$) among samples of different thicknesses, and the actual slope is independent of the external magnetic field. The external magnetic field-dependent $\Delta E$ at different cavity mode resonances carries information about the internal electronic states, which is helpful for further understanding the energy band information of $FePS_3$. The more pronounced $\Delta E$ changes with the magnetic field, the larger the magnetic moment of the electron. Regarding $B_0$, as shown in Figure 3d, the magnitude of $B_0$ increases with thickness. The increase in $B_0$ might be because as the $FePS_3$ gets thicker, the degree of inversion symmetry breaking between $FePS_3$ and substrate increases. When the thickness is fixed, for thin samples (S2 and S3), the difference in $B_0$ between different wavelengths is small, while for thick samples (S1, 1500 nm), the magnitude of $B_0$ varies greatly with wavelength ranging from about -0.27 T to -0.14 T. Our results suggest that the cavity-enhanced CD technique is an effective probe to detect the broken time-reversal symmetry, inversion asymmetry-induced interfacial spin-orbit magnetic field, and hidden interface state.

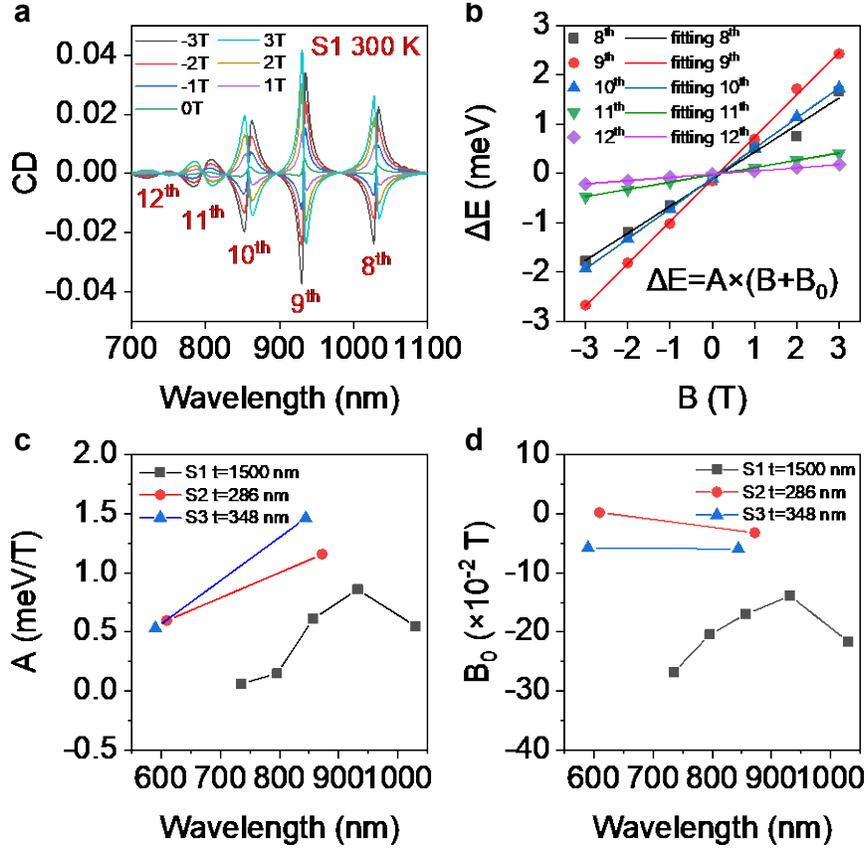

**Figure 3.** Magnetic field dependence of cavity-enhanced CD spectra of FePS$_3$ at 300 K. (a) Magnetic field-dependent CD spectra of S1 flake (1500 nm). Different CD resonances are marked with their order number ($p$). (b) Magnetic field-dependent energy splitting ($\Delta E$) at specific wavelengths, extracted from (a). The symbols represent the experimental results, and the lines are the corresponding linear fitting results using the formula $\Delta E = A \times (B + B_0)$. Linear fitting parameters (c) $A$ and (d) $B_0$ for three samples of different thicknesses.

To explore the relationships between the nonzero CD at 0 T and different magnetic states of FePS$_3$, we conducted temperature-dependent CD spectra measurements at 0 T. The intensity contour plot of the temperature-dependent CD is displayed in Figure 4a, which shows that the CD magnitude is very weak in the PM state ($T>T_N$), while the CD value increases dramatically below the magnetic phase transition temperature ($T_N \approx 118$ K), reaching a magnitude comparable to that of the CD at -3 T for the PM FePS$_3$. Below $T_N$, the AFM FePS$_3$, which lacks net magnetization, retains the monoclinic structure. As shown in Figure 1b, the combined centrosymmetric structure and zig-zag AFM ordering preserve the inversion symmetry within FePS$_3$ itself, which would prohibit the NCD. However, similar to the case of PM FePS$_3$, the broken inversion symmetry at the interface can induce a finite NCD at 0 T. Besides the structural inversion asymmetry, the spin-related inversion asymmetry between the zig-zag AFM FePS$_3$ and the nonmagnetic substrate also significantly contributes to the larger CD below $T_N$. Before and after the magnetic phase transition, in addition to the obvious difference in CD magnitude, the peak position and linewidth of the CD also vary, as shown in Figure 4b, which results from the change in the dielectric tensor below and above $T_N$. To investigate the applied magnetic field response of CD in the AFM state, we measured the CD spectra of the S1 flake (1500 nm) under different external magnetic fields at 65 K, as shown in Figure 4c. For the AFM FePS$_3$, the CD and, thus, the $\Delta E$ is almost independent of the external magnetic field below 3 T, in contrast to the linear external magnetic field dependence of $\Delta E$ in the PM phase. Since the critical spin-flop field for FePS$_3$ is 35 T,[36] an out-of-plane magnetic field of 3 T is insufficient to rotate the spins. Therefore, the spins in FePS$_3$ remain in the zig-zag AFM order, and the CD, which couples to the combined spin and structural inversion asymmetry, would not shift as a function of the external magnetic field below 3 T. Our results demonstrate that constructing an asymmetry interface enables CD to be an effective mean for investigating the magnetic properties of fully compensated collinear AFM materials, including tracking the magnetic phase transition temperature and the spin-flop field.

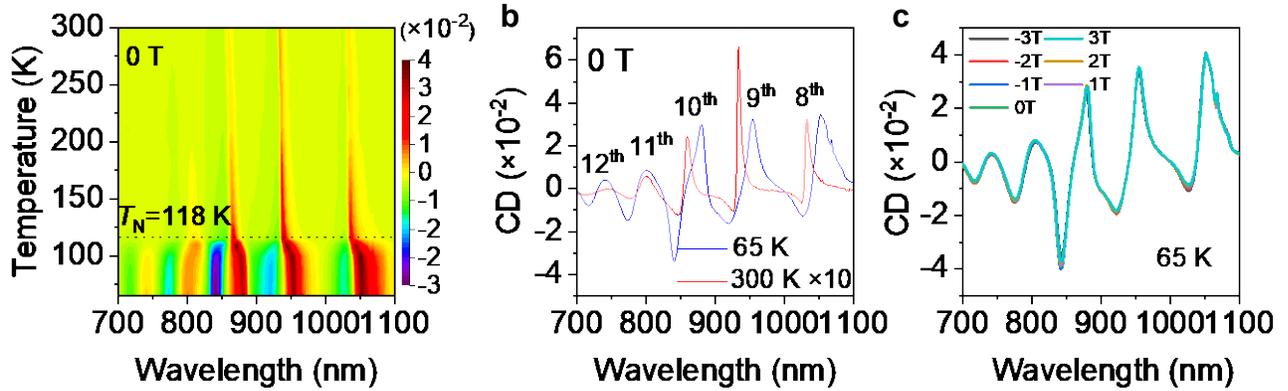

**Figure 4.** Temperature dependence of cavity-enhanced CD spectra of FePS$_3$. (a) Intensity contour plot of temperature-dependent CD spectra at 0 T. (b) Comparison of CD spectra for FePS$_3$ at 65 K and 300 K. The one at 300 K is multiplied by 10. Different CD resonances are marked with their order number ($p$). (c) Magnetic field-dependent CD spectra of FePS$_3$ at 65 K.

In conclusion, we demonstrated that cavity-enhanced CD spectroscopy is an effective tool for probing the magnetic order, time-reversal symmetry breaking, and interfacial asymmetries in FePS$_3$. We showed that the external magnetic field induces MCD in the PM state due to the broken time-reversal symmetry, which can be significantly enhanced through the interference effect. Through the creation of an asymmetric FePS$_3$/substrate interface, we induced a weak but detectable CD signal that originates from the interfacial spin-orbit magnetic field ($B_0$), a feature not accessible by traditional magnetic probes, like Gauss meter, magnetic force microscopy, and magnetization measurements. In the AFM state, the CD is strongly coupled with the intrinsic zig-zag AFM spin alignment, providing a straightforward way to measure the magnetic phase transition temperature and the spin-flop field, overcoming the challenges of detecting antiferromagnetism due to the lack of net magnetization in fully compensated antiferromagnets. Our work establishes cavity-enhanced CD as a highly sensitive method for detecting broken symmetries in AFM materials and opens the door to using this technique in the study of hidden spin polarizations and interfaces. Furthermore, the ability to tune the CD signal spectrally and in magnitude through external stimuli, such as magnetic fields, temperature, and strain, highlights the potential of this approach in developing advanced nanophotonic devices, particularly for circular dichroism-based optical components.

ASSOCIATED CONTENT

**Supporting Information**.

The Supporting Information is available free of charge.

> Cavity-enhanced effect on optical activity and magnetooptical effect, sample preparation, experimental and simulated Reflectance and CD with different thickness of $FePS_3$, distributed Bragg reflector (DBR) substrate-enhanced CD of thin $FePS_3$, CD spectra of S1 sample at +0 T and -0 T, and full magnetic field-dependent CD spectra of $FePS_3$ (sample S2 and S3)

AUTHOR INFORMATION

**Corresponding Authors**

Jun Zhang - *State Key Laboratory of Superlattices and Microstructures, Institute of Semiconductors, Chinese Academy of Sciences, Beijing 100083, China; College of Materials Science and Opto-Electronic Technology, University of Chinese Academy of Sciences, Beijing 100049, China;* Email: zhangjwill@semi.ac.cn

Chao Shen - *State Key Laboratory of Superlattices and Microstructures, Institute of Semiconductors, Chinese Academy of Sciences, Beijing 100083, China; College of Materials*

*Science and Opto-Electronic Technology, University of Chinese Academy of Sciences, Beijing 100049, China;* Email: shenchao@semi.ac.cn

**Authors**

**Shu-Liang Ren** - *State Key Laboratory of Superlattices and Microstructures, Institute of Semiconductors, Chinese Academy of Sciences, Beijing 100083, China; College of Materials Science and Opto-Electronic Technology, University of Chinese Academy of Sciences, Beijing 100049, China*

**Simin Pang** - *State Key Laboratory of Superlattices and Microstructures, Institute of Semiconductors, Chinese Academy of Sciences, Beijing 100083, China; College of Materials Science and Opto-Electronic Technology, University of Chinese Academy of Sciences, Beijing 100049, China*

**Shan Guan** - *State Key Laboratory of Superlattices and Microstructures, Institute of Semiconductors, Chinese Academy of Sciences, Beijing 100083, China*

**Yu-Jia Sun** - *State Key Laboratory of Superlattices and Microstructures, Institute of Semiconductors, Chinese Academy of Sciences, Beijing 100083, China; College of Materials Science and Opto-Electronic Technology, University of Chinese Academy of Sciences, Beijing 100049, China*

**Tian-Yu Zhang** - *College of Materials Science and Opto-Electronic Technology, University of Chinese Academy of Sciences, Beijing 100049, China*

**Nai Jiang** - *State Key Laboratory of Superlattices and Microstructures, Institute of Semiconductors, Chinese Academy of Sciences, Beijing 100083, China; College of Materials*

*Science and Opto-Electronic Technology, University of Chinese Academy of Sciences, Beijing 100049, China*

**Jiaqi Guo** - *State Key Laboratory of Superlattices and Microstructures, Institute of Semiconductors, Chinese Academy of Sciences, Beijing 100083, China*

**Hou-Zhi Zheng** - *State Key Laboratory of Superlattices and Microstructures, Institute of Semiconductors, Chinese Academy of Sciences, Beijing 100083, China; College of Materials Science and Opto-Electronic Technology, University of Chinese Academy of Sciences, Beijing 100049, China*

**Jun-Wei Luo** - *State Key Laboratory of Superlattices and Microstructures, Institute of Semiconductors, Chinese Academy of Sciences, Beijing 100083, China; College of Materials Science and Opto-Electronic Technology, University of Chinese Academy of Sciences, Beijing 100049, China*

**Ping-Heng Tan** - *State Key Laboratory of Superlattices and Microstructures, Institute of Semiconductors, Chinese Academy of Sciences, Beijing 100083, China; College of Materials Science and Opto-Electronic Technology, University of Chinese Academy of Sciences, Beijing 100049, China*

**Author Contributions**

J.Z. conceived the project, C.S. and J.Z. designed the experiments. S.-L.R., S.P., and N. J. conducted experiments. T.-Y.Z. and S.-L.R. conducted the simulations of reflectance and CD spectra. S.G. and J.L. analyzed the interfacial spin-orbit magnetic field. S.-L.R., S.P., and J.Z. analyzed the data and wrote the manuscript. All the authors discussed the results and revised the paper. †These authors contributed equally.

**Notes**

The authors declare no competing financial interest.

ACKNOWLEDGMENTS

J. Z. acknowledges the CAS Project for Young Scientists in Basic Research (YSBR-120), National Natural Science Foundation of China (12074371), Research Equipment Development Project of Chinese Academy of Sciences (YJKYYQ20210001). C.S. acknowledges the support from the Youth Innovation Promotion Association, Chinese Academy of Sciences (2019114).

REFERENCES

(1) Sun, Z. Y.; Yi, Y. F.; Song, T. C.; Clark, G.; Huang, B.; Shan, Y. W.; Wu, S.; Huang, D.; Gao, C. L.; Chen, Z. H.; et al. Giant nonreciprocal second-harmonic generation from antiferromagnetic bilayer $CrI_3$. *Nature* **2019,** *572*, 497-501.

(2) McCreary, A.; Mai, T. T.; Utermohlen, F. G.; Simpson, J. R.; Garrity, K. F.; Feng, X. Z.; Shcherbakov, D.; Zhu, Y. L.; Hu, J.; Weber, D.; et al. Distinct magneto-Raman signatures of spin-flip phase transitions in $CrI_3$. *Nature Commun.* **2020,** *11*, 3879.

(3) Lee, K.; Dismukes, A. H.; Telford, E. J.; Wiscons, R. A.; Wang, J.; Xu, X. D.; Nuckolls, C.; Dean, C. R.; Roy, X.; Zhu, X. Y. Magnetic Order and Symmetry in the 2D Semiconductor CrSBr. *Nano Lett.* **2021,** *21*, 3511-3517.

(4) Gong, C.; Zhang, X. Two-dimensional magnetic crystals and emergent heterostructure devices. *Science* **2019,** *363*, eaav4450.

(5) Xing, S. C.; Zhou, J.; Zhang, X. G.; Elliott, S.; Sun, Z. M. Theory, properties and engineering of 2D magnetic materials. *Prog. Mater. Sci.* **2023,** *132*, 101036.

(6) Guo, Y. L.; Wang, B.; Zhang, X. W.; Yuan, S. J.; Ma, L.; Wang, J. L. Magnetic two-dimensional layered crystals meet with ferromagnetic semiconductors. *InfoMat* **2020,** *2*, 639-655.

(7) Kurebayashi, H.; Garcia, J. H.; Khan, S.; Sinova, J.; Roche, S. Magnetism, symmetry and spin transport in van der Waals layered systems. *Nat. Rev. Phys.* **2022,** *4*, 150-166.

(8) Stephens, P. J. Magnetic circular-dichroism. *Annu. Rev. Phys. Chem.* **1974,** *25*, 201-232.

(9) Woody, R. W. Circular-Dichroism. In *Biochemical Spectroscopy*; Academic Press, 1995; pp 34-71.

(10) Govorov, A. O.; Fan, Z. Y.; Hernandez, P.; Slocik, J. M.; Naik, R. R. Theory of circular dichroism of nanomaterials comprising chiral molecules and nanocrystals: plasmon enhancement, dipole interactions, and dielectric effects. *Nano Lett.* **2010,** *10*, 1374-1382.

(11) Han, B.; Gao, X. Q.; Lv, J. W.; Tang, Z. Y. Magnetic circular dichroism in nanomaterials: new opportunity in understanding and modulation of excitonic and plasmonic resonances. *Adv. Mater.* **2020,** *32*, 1801491

(12) Mason, W. R. Introduction. In *A Practical Guide to Magnetic Circular Dichroism Spectroscopy*; John Wiley & Sons, Inc., 2007; pp 1-3.

(13) Barron, L. D., *Molecular Light Scattering and Optical Activity*; Cambridge University Press, 2009.

(14) Mak, K. F.; Shan, J.; Ralph, D. C. Probing and controlling magnetic states in 2D layered magnetic materials. *Nat. Rev. Phys.* **2019**, *1*, 646-661.

(15) Jungwirth, T.; Marti, X.; Wadley, P.; Wunderlich, J. Antiferromagnetic spintronics. *Nat. Nanotechnol.* **2016**, *11*, 231-241.

(16) Duine, R. A.; Lee, K. J.; Parkin, S. S. P.; Stiles, M. D. Synthetic antiferromagnetic spintronics. *Nat. Phys.* **2018**, *14*, 217-219.

(17) Nemec, P.; Fiebig, M.; Kampfrath, T.; Kimel, A. V. Antiferromagnetic opto-spintronics. *Nat. Phys.* **2018**, *14*, 229-241.

(18) Lee, J. U.; Lee, S.; Ryoo, J. H.; Kang, S.; Kim, T. Y.; Kim, P.; Park, C. H.; Park, J. G.; Cheong, H. Ising-type magnetic ordering in atomically thin $FePS_3$. *Nano Lett.* **2016**, *16*, 7433-7438.

(19) Lançon, D.; Walker, H. C.; Ressouche, E.; Ouladdiaf, B.; Rule, K. C.; McIntyre, G. J.; Hicks, T. J.; Ronnow, H. M.; Wildes, A. R. Magnetic structure and magnon dynamics of the quasi-two-dimensional antiferromagnet $FePS_3$. *Phys. Rev. B* **2016**, *94*, 214407.

(20) Lee, Y.; Son, S.; Kim, C.; Kang, S.; Shen, J. Y.; Kenzelmann, M.; Delley, B.; Savchenko, T.; Parchenko, S.; Na, W.; et al. Giant magnetic anisotropy in the atomically thin van der Waals antiferromagnet $FePS_3$. *Adv. Electron. Mater.* **2023**, *9*, 2200650.

(21) Zhang, H. Q.; Ni, Z. L.; Stevens, C. E.; Bai, A. F.; Peiris, F.; Hendrickson, J. R.; Wu, L.; Jariwala, D. Cavity-enhanced linear dichroism in a van der Waals antiferromagnet. *Nat. Photonics* **2022**, *16*, 311-317.

(22) Ni, Z. L.; Huang, N.; Haglund, A. V.; Mandrus, D. G.; Wu, L. Observation of giant surface second-harmonic generation coupled to nematic orders in the van der Waals antiferromagnet $FePS_3$. *Nano Lett.* **2022**, *22*, 3283-3288.

(23) Zhang, X. X.; Jiang, S. W.; Lee, J.; Lee, C. G.; Mak, K. F.; Shan, J. Spin Dynamics slowdown near the antiferromagnetic critical point in atomically thin $FePS_3$. *Nano Lett.* **2021**, *21*, 5045-5052.

(24) Liu, S.; del Aguila, A. G.; Bhowmick, D.; Gan, C. K.; Do, T. T. H.; Prosnikov, M. A.; Sedmidubsky, D.; Sofer, Z.; Christianen, P. C. M.; Sengupta, P.; et al. Direct observation of magnon-phonon strong coupling in two-dimensional antiferromagnet at high magnetic fields. *Phys. Rev. Lett.* **2021**, *127*, 097401.

(25) Du, K. Z.; Wang, X. Z.; Liu, Y.; Hu, P.; Utama, M. I. B.; Gan, C. K.; Xiong, Q. H.; Kloc, C. Weak van der Waals stacking, wide-range band gap, and Raman study on ultrathin layers of metal phosphorus trichalcogenides. *Acs Nano* **2016**, *10*, 1738-1743.

(26) Jernberg, P.; Bjarman, S.; Wappling, R. $FePS_3$: A first-order phase transition in a "2D" Ising antiferromagnet. *J. Magn. Magn. Mater.* **1984**, *46*, 178-190.

(27) Zhou, F. R.; Liu, H. H.; Zajac, M.; Hwangbo, K.; Jiang, Q. N.; Chu, J. H.; Xu, X. D.; Arslan, I.; Gage, T. E.; Wen, H. D. Ultrafast nanoimaging of spin-mediated shear waves in an acoustic cavity. *Nano Lett.* **2023**, *23*, 10213-10220.

(28) Feng, W. X.; Guo, G. Y.; Zhou, J.; Yao, Y. G.; Niu, Q. Large magneto-optical Kerr effect in noncollinear antiferromagnets $Mn_3X$ (X = Rh, Ir, Pt). *Phys. Rev. B* **2015**, *92*, 144426.

(29) Kats, M. A.; Blanchard, R.; Genevet, P.; Capasso, F. Nanometre optical coatings based on strong interference effects in highly absorbing media. *Nat. Mater.* **2013**, *12*, 20-24.

(30) Wu, Y. J.; Shen, C.; Tan, Q. H.; Shi, J.; Liu, X. F.; Wu, Z. H.; Zhang, J.; Tan, P. H.; Zheng, H. Z. Valley Zeeman splitting of monolayer $MoS_2$ probed by low-field magnetic circular dichroism spectroscopy at room temperature. *Appl. Phys. Lett.* **2018,** *112*, 153105.

(31) Mason, W. R. Theoretical Framework: Definition of MCD Terms. In *A Practical Guide to Magnetic Circular Dichroism Spectroscopy*; John Wiley & Sons, Inc., 2007; pp 14-35.

(32) Sugano, S.; Kojima, N. *Magneto-Optics*; Springer Science & Business Media, 2013.

(33) Chen, L.; Gmitra, M.; Vogel, M.; Islinger, R.; Kronseder, M.; Schuh, D.; Bougeard, D.; Fabian, J.; Weiss, D.; Back, C. H. Electric-field control of interfacial spin-orbit fields. *Nat. Electron.* **2018,** *1*, 350-355.

(34) Soumyanarayanan, A.; Reyren, N.; Fert, A.; Panagopoulos, C. Emergent phenomena induced by spin-orbit coupling at surfaces and interfaces. *Nature* **2016,** *539*, 509-517.

(35) Ruan, Y. L.; Jarvis, R. A.; Rode, A. V.; Madden, S.; Luther-Davies, B. Wavelength dispersion of Verdet constants in chalcogenide glasses for magneto-optical waveguide devices. *Opt. Commun.* **2005,** *252*, 39-45.

(36) Wildes, A. R.; Lançon, D.; Chan, M. K.; Weickert, F.; Harrison, N.; Simonet, V.; Zhitomirsky, M. E.; Gvozdikova, M. V.; Ziman, T.; Ronnow, H. M. High field magnetization of $FePS_3$. *Phys. Rev. B* **2020,** *101*, 024415.

# Supporting Information for "Cavity-enhanced circular dichroism in a van der Waals antiferromagnet"

Shu-Liang Ren[1,2,†], Simin Pang[1,2,†], Shan Guan[1], Yu-Jia Sun[1,2], Tian-Yu Zhang[2], Nai Jiang[1,2], Jiaqi Guo[1], Hou-Zhi Zheng[1,2], Jun-Wei Luo[1,2], Ping-Heng Tan[1,2], Chao Shen[1,2]*, and Jun Zhang[1,2]*

[1]State Key Laboratory of Superlattices and Microstructures, Institute of Semiconductors, Chinese Academy of Sciences, Beijing 100083, China.

[2]College of Materials Science and Opto-Electronic Technology, University of Chinese Academy of Sciences, Beijing 100049, China.

*Corresponding authors. Email: shenchao@semi.ac.cn (S.C.); zhangjwill@semi.ac.cn (J.Z.)

[†]These authors contributed equally to this work.

**Section 1. Cavity-enhanced effect on optical activity and magnetooptical effect**

Natural circular dichroism (NCD) and optical rotation do not change sign under the time-reversal symmetry, opposite to the cases for magnetic circular dichroism (MCD) and Faraday rotation. Therefore, as shown in Figure S1a, assuming that the absorption of RCP light is more significant than that of LCP light, i.e., CD < 0, the reflected lights maintaining their chirality pass through the sample again, and the NCD will further accumulate. However, since the MCD would change sign when the propagation direction of light is reversed, the MCD accumulated in the forward propagation would be compensated by the backward one (Figure S1b). Differently, the polarization angle changes are additive for the forward and backward propagating light in the case of Faraday rotation (Figure S1d), while compensated in the case of optical rotation (Figure S1c).[1]

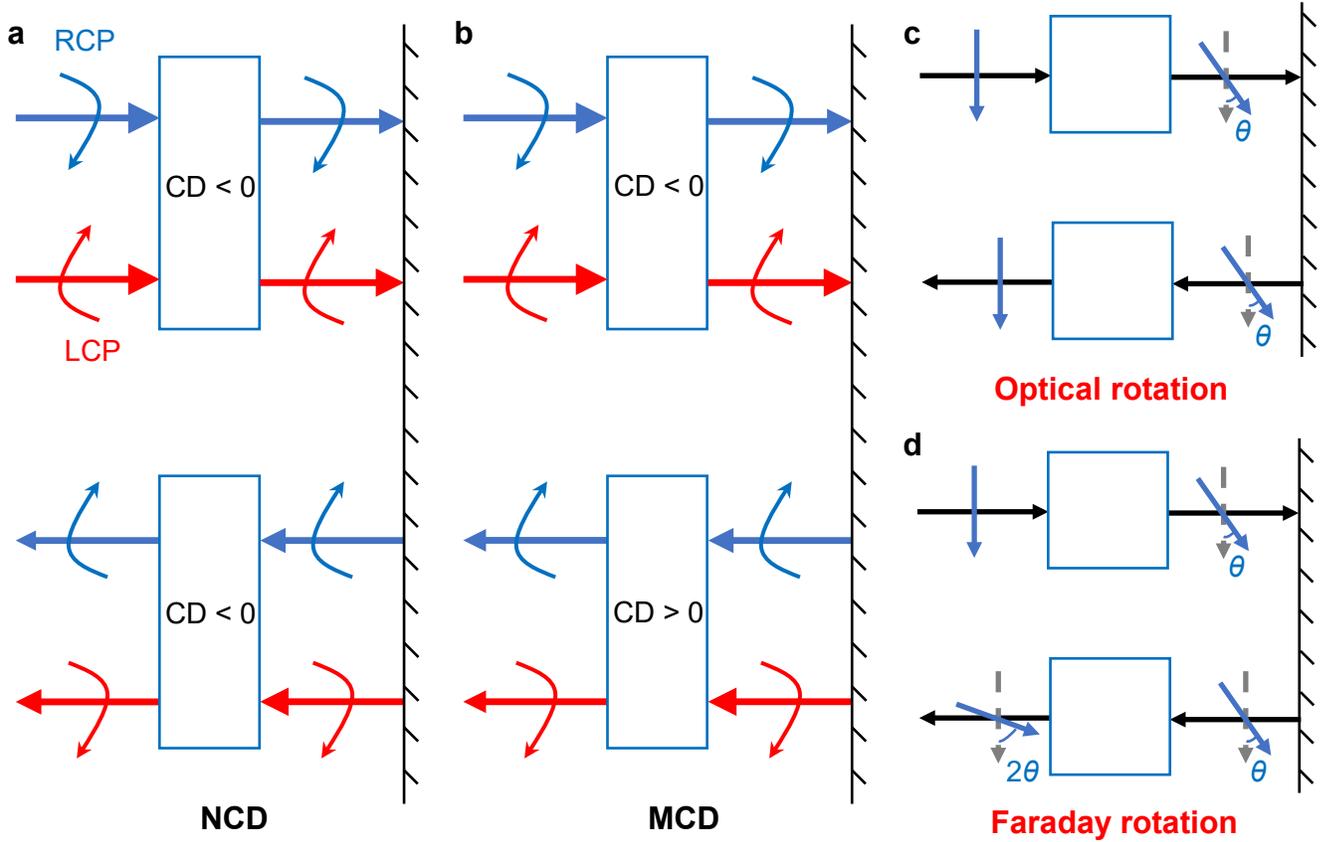

**Figure S1.** Cavity-enhanced effect on optical activity and magnetooptical effect. (a) NCD. (b) MCD. The blue and red curved arrows show the chirality of the RCP and LCP lights, respectively. (c) Optical rotation. (d) Faraday rotation. The blue solid arrows denote the actual polarization directions of the propagating light. The gray dotted arrows indicate the initial polarization directions for comparison. $\theta$ and $2\theta$ represent the polarization angle changes.

**Section 2. Sample preparation**

The FePS$_3$ flakes were mechanically exfoliated on 90 nm SiO$_2$/Si and DBR substrates in air and then transferred into the vacuum chamber of a microscopic cryostat. In our experiment, we did not observe any significant degradation of the FePS$_3$ flakes in the air. The flake thickness was determined using a step profiler.

**Section 3. Reflectance CD measurements**

As shown in Figure 1d, the CD measurements were performed in a home-built microscopic snystem. Light emitted from a supercontinuum white light source is monochromatized by a monochromator (Horiba Jobin-Yvon iHR320), producing excitation light with a tunable wavelength. A chopper (Stanford Research System, SR540) with a modulation frequency of 177 Hz is used to modulate the intensity of the incidet light. After passing through a linear polarizer (LP), the linearly polarized light is directed to the photoelastic modulator (PEM, Hinds Instruments), polarized at 45° to the fast axis of the PEM. The PEM periodically modulates the linearly polarized light into LCP and RCP light for CD measurements, at a frequency of 50 kHz. After phase modulation, the light is incident vertically on the sample through a 10x objective with a numerical aperture NA=0.25. The light reflected from the sample is collected by a Si photodetector. Using two lock-in amplifiers (Stanford Research System, SR830), one tuned to the chopper frequency (177 Hz) and one tuned to the PEM frequency (50 kHz), the reflectance and CD spectra are obtained simultaneously. The chopper-synchronized signal corresponds to the average of the reflectance of LCP and RCP light $[(R_+ + R_-)/2]$, while the PEM-synchronized signal corresponds to the difference $(R_+ - R_-)$. A superconducting magnet provides a variable out-of-plane magnetic field along the incident direction of the excitation light ($z$-axis) for the MCD measurements. For all temperature-dependent measurements, samples were measured in a microscopic vacuum cryostat (65 to 300 K) in a high vacuum.

**Section 4. Reflectance and CD simulations**

For simplicity, the dielectric tensor of the monoclinic $FePS_3$ can be described as:[2-4]

$$\varepsilon = \begin{pmatrix} \varepsilon_{xx} & -iQ_v m_3 & 0 \\ iQ_v m_3 & \varepsilon_{xx} & 0 \\ 0 & 0 & \varepsilon_{zz} \end{pmatrix}, \tag{1}$$

where $Q_v$ is Voigt constant and roughly proportional to the magnitude of the magnetic field, and $m_3$ is the unit vector of the out-of-plane component of the magnetization. This form of the dielectric tensor is valid for the magnetic medium assuming that (1) at least a threefold rotation axis along the z-axis exists and (2) the polar case is satisfied with the magnetization being parallel to the z-direction (out-of-plane direction). Although the bulk FePS$_3$ is monoclinic, the monolayer has $D_{3d}$ symmetry which includes a threefold rotation axis (z-axis),[5,6] satisfying assumption (1). Additionally, FePS$_3$ exhibits Ising-type antiferromagnetic ordering,[7] satisfying assumption (2).

Combining the dielectric tensor (eq 1) and Maxwell's equations, the eigenmodes of the light waves propagating in FePS$_3$ can be rigorously derived. Considering that the light incidence is parallel to the z-axis (normal incidence), i.e., polar effect, solving the normal modes, the eigenvalues can be obtained[8,9]

$$\tilde{n}_\pm \approx \sqrt{\varepsilon_{xx}}\left(1 \pm \frac{Q_v}{2}\right), \tag{2}$$

where the complex refractive index $\tilde{n}_+$ ($\tilde{n}_-$) denotes the eigenvalue of left- (right-) circularly polarized waves, respectively. For eq 2, it is assumed that the non-diagonal term of the dielectric tensor (eq 1) is smaller than the diagonal term ($Q_v \ll \varepsilon_{xx}$), which is valid for FePS$_3$. The average complex refractive index ($\bar{n}$) and complex refractive index difference ($\Delta\tilde{n}$) can be written as

$$\begin{cases} \bar{n} = \frac{1}{2}(\tilde{n}_+ + \tilde{n}_-) = \sqrt{\varepsilon_{xx}} \\ \Delta\tilde{n} = \tilde{n}_+ - \tilde{n}_- = \bar{n}Q_v \end{cases}. \tag{3}$$

According to Fresnel's equations, the reflection coefficients of LCP and RCP waves are[8,9]

$$r(\tilde{n}_\pm) = \frac{1-\tilde{n}_\pm}{1+\tilde{n}_\pm}. \tag{4}$$

Here $r(\tilde{n}_+) = r_x + ir_y$ and $r(\tilde{n}_-) = r_x - ir_y$, from which we obtain

$$\begin{cases} r_x = \frac{1}{2}[r(\tilde{n}_+) + r(\tilde{n}_-)] \\ r_y = \frac{1}{2}i[r(\tilde{n}_+) - r(\tilde{n}_-)] \end{cases}. \tag{5}$$

With the help of eq 2-4 and ignoring terms of higher order in $Q_v$, eq 5 can be rewritten as

$$\begin{cases} r_x = \frac{1-\bar{n}}{1+\bar{n}} \\ r_y = \frac{-i\Delta\tilde{n}}{(1+\bar{n})^2} \end{cases}. \tag{6}$$

Therefore, eq 4 can also be written as

$$r(\tilde{n}_\pm) = \frac{1-\bar{n}}{1+\bar{n}} \pm \frac{\Delta\tilde{n}}{(1+\bar{n})^2}, \tag{7}$$

according to which we simulate the reflection coefficients of LCP and RCP waves, mainly considering the complex refractive index difference ($\Delta\tilde{n}$) which is proportional to the Voigt constant $Q_v$. The total reflectance of LCP and RCP waves can be calculated by $R(\tilde{n}_\pm) = r(\tilde{n}_\pm)^* \cdot r(\tilde{n}_\pm)$.

The peaks in the reflectance spectrum can be described as a Gaussian line centered at energy $E_0$, $R(E) = R_0 \exp\left[-\frac{(E-E_0)^2}{2w^2}\right]$. Under an external perpendicular magnetic field, the Zeeman energy splitting ($\Delta E$) occurs with two components centered at $E_0 \pm \frac{\Delta E}{2}$. In most cases, $\Delta E$ is much smaller than the peak linewidth ($\Delta E < 0.1w$).[10] The rigid-shift approximation[11] assuming that the band shifts due to the Zeeman effect but does not change shape can be applied here. Therefore, the reflectance spectra corresponding to LCP and RCP lights can be expressed by

$$\begin{cases} R_+(E) = R_0 exp\left[-\frac{\left[E-\left(E_0-\frac{\Delta E}{2}\right)\right]^2}{2w^2}\right] = R\left(E+\frac{\Delta E}{2}\right) \\ R_-(E) = R_0 exp\left[-\frac{\left[E-\left(E_0+\frac{\Delta E}{2}\right)\right]^2}{2w^2}\right] = R\left(E-\frac{\Delta E}{2}\right) \end{cases}. \quad (8)$$

We defined that the average reflectance without polarization (the signal synchronized with the chopper) $R = [R_+(E) + R_-(E)]/2$. The CD can be expressed by

$$CD = \frac{R_+(E)-R_-(E)}{[R_+(E)+R_-(E)]/2} \cong \frac{1}{R}\frac{dR(E)}{dE}\Delta E. \quad (9)$$

Therefore, the CD can be calculated using eq 9. The refractive index difference ($\Delta n$) between LCP and RCP lights is related to $\Delta E$,

$$\Delta n = C \cdot p^2 \cdot \Delta E \cdot f'(\omega_0), \quad (10)$$

where $C$ is a constant independent of the light frequency, $p^2$ is transition strength, and $f'(\omega_0)$ is a function related to the energy of absorption.[12] As shown in Figure 1e, we assumed that $f'(\omega_0)$ can be approximated as a constant here, since $k$ is weak in the wavelength region we investigated. Therefore, we can consider that $CD \propto \frac{1}{R}\frac{dR(E)}{dE}\Delta n$.

**Section 5. Experimental and simulated reflectance and CD spectra of FePS$_3$ with different thicknesses**

As shown in Figure S2a, the 29 nm thick sample on 90 nm SiO$_2$/Si substrate is unable to support cavity modes in the detection wavelength range, and no CD signal is observed.

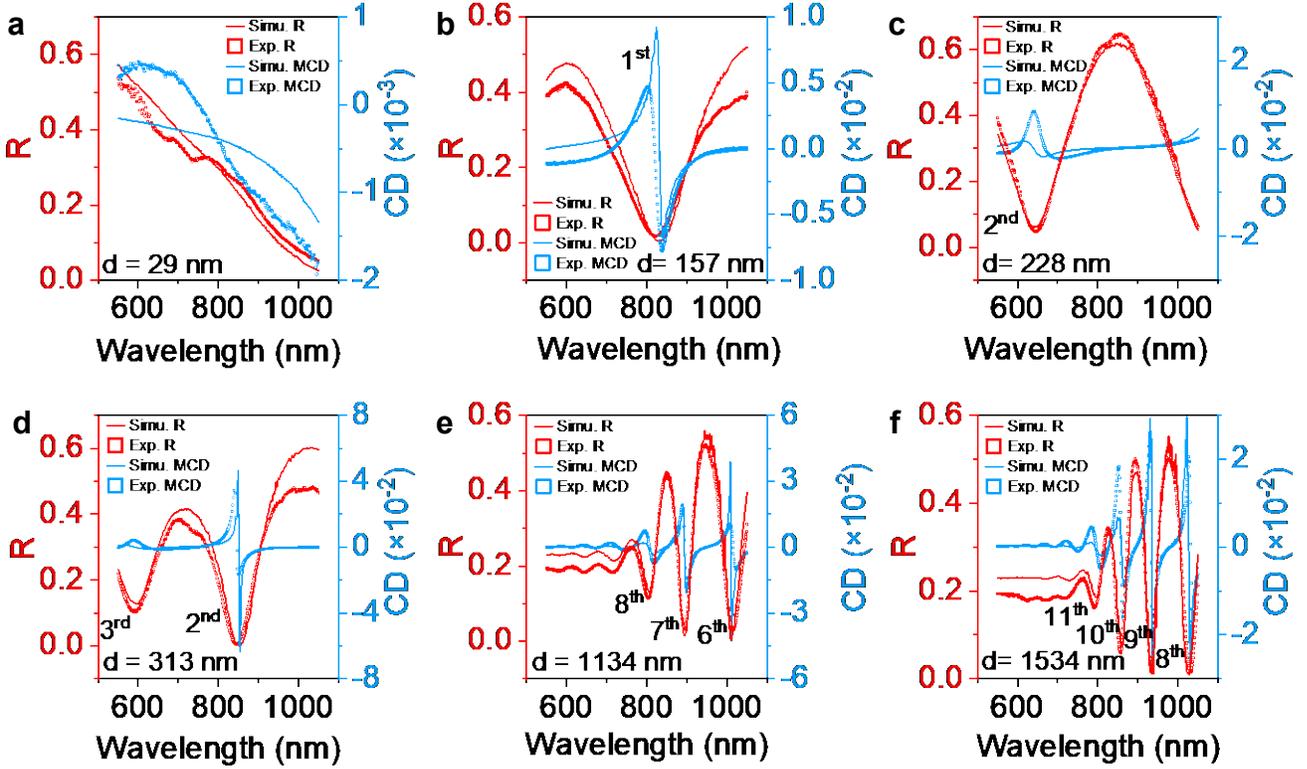

**Figure S2.** Experimental reflectance (red) and CD (blue) spectra of FePS$_3$ with different thicknesses on 90 nm SiO$_2$/Si substrate at -3 T and 300 K, together with the simulated curves. (a) 29 nm. (b) 157 nm. (c) 228 nm. (d) 313 nm. (e) 1134 nm. (f) 1534 nm. Different CD resonances are marked with their order number ($p$).

## Section 6. Using a distributed Bragg reflector (DBR) substrate to enhance the CD of thin FePS$_3$

As shown in Figure S3a, the distributed Bragg reflector (DBR) substrate consists of multi-layer alternating dielectric pairs (SiO$_2$ and Ta$_2$O$_5$). The thickness of SiO$_2$ (low refractive index) is 130.14 nm and that of Ta$_2$O$_5$ (high refractive index) is 92.68 nm. Since the structure of DBR is chiral, nonzero CD exists at 0 T, which is independent of the external magnetic field below 3 T, as shown in Figure S3b. From Figure S3c, we can see that using the DBR substrate can enhance the CD for the thin FePS$_3$ sample (80 nm here). In addition, the energy splitting ($\Delta E$) of 80 nm FePS$_3$ on DBR also varies linearly with the magnetic field (Figure S3d).

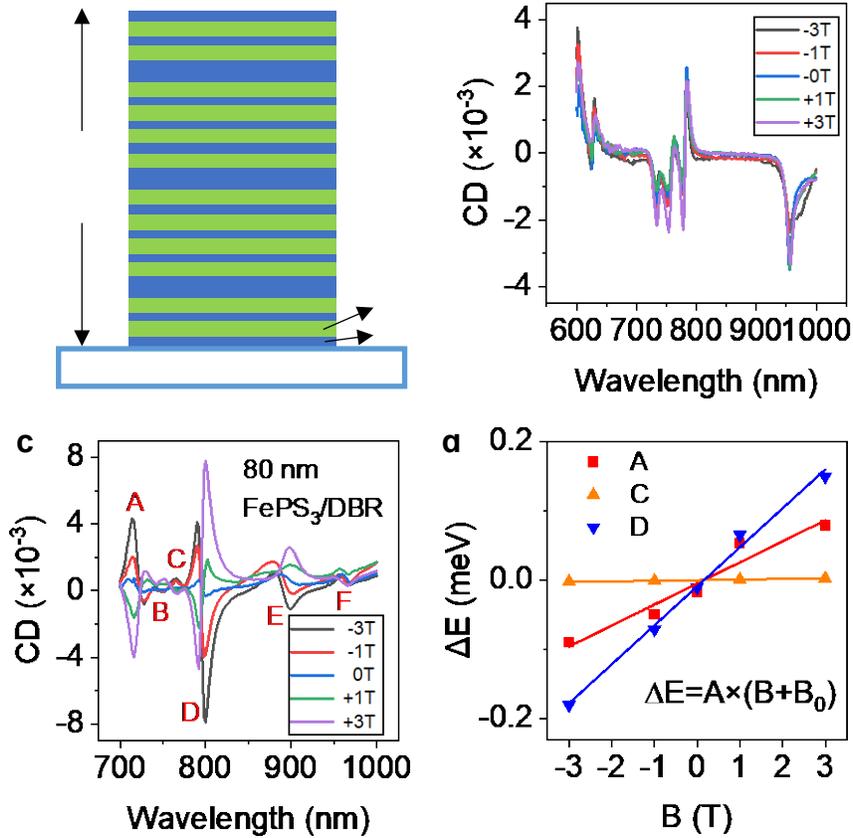

**Figure S3.** CD spectra of DBR substrate and 80 nm FePS$_3$ on DBR at 285 K. (a) Schematic diagram of the structure of DBR formed by periodic arrangements of SiO$_2$ and Ta$_2$O$_5$. The 2H represents twice the corresponding thickness. Magnetic field-dependent CD spectra of (b) DBR substrate and (c) 80 nm FePS$_3$ on DBR substrate. (d) Magnetic field-dependent energy splitting ($\Delta E$) at specific wavelengths, extracted from (c). The symbols represent the experimental results, and the lines are the corresponding linear fitting results using the formula $\Delta E = A \times (B + B_0)$.

**Section 7. Examining whether FePS$_3$ shows ferromagnetism at room temperature**

We defined the zero magnetic field falling from the positive magnetic field as +0 T, and the zero magnetic field arising from the negative magnetic field as -0 T. At room temperature, the CD at +0 T and -0 T are the same, as shown in Figure S4, indicating that the nonzero magnetic field is not due to intrinsic ferromagnetism but to interface structural asymmetry.

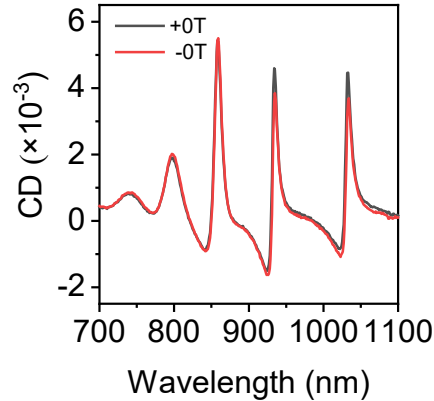

**Figure S4.** CD spectra measurements of the S1 sample at +0 T and -0 T at room temperature. +0 T is the magnetic field sweeping from the +3 T to -3 T, while -0 T corresponds to the opposite sweeping direction.

**Section 8. Full magnetic field-dependent CD spectra of FePS$_3$ (sample S2 and S3) at room temperature**

As shown in Figure S5, the magnetic field-dependent CD spectra measured at room temperature for the FePS$_3$ samples S2 (286 nm) and S3 (348 nm) reflect that the wavelengths corresponding to the extreme values of the CD depend on the sample thickness.

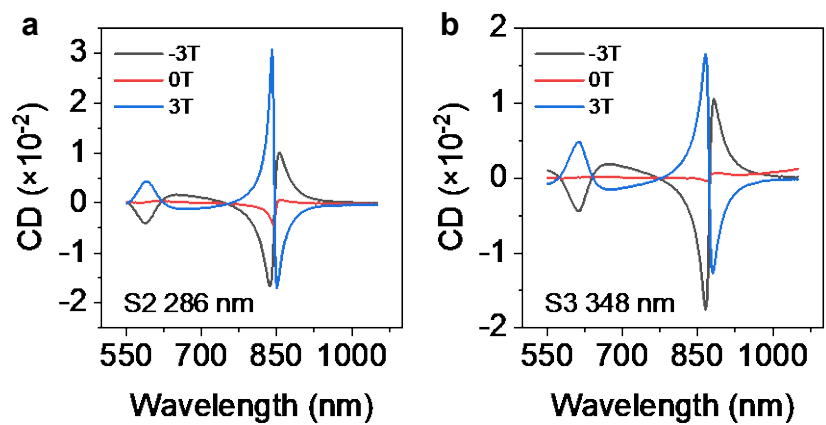

**Figure S5.** Magnetic field-dependent CD spectra measured at room temperature for FePS$_3$ with different thicknesses. (a) 286 nm. (b) 348 nm.

REFERENCES

(1) Barron, L. D., *Molecular Light Scattering and Optical Activity*; Cambridge University Press, 2009.
(2) Ebert, H. Magneto-optical effects in transition metal systems. *Rep. Prog. Phys.* **1996,** *59*, 1665-1735.
(3) Erskine, J. L.; Stern, E. A. Calculation of the M23 magneto-optical absorption spectrum of ferromagnetic nickel. *Phys. Rev. B* **1975,** *12*, 5016-5024.
(4) Huang, B.; Clark, G.; Navarro-Moratalla, E.; Klein, D. R.; Cheng, R.; Seyler, K. L.; Zhong, D.; Schmidgall, E.; McGuire, M. A.; Cobden, D. H.; et al. Layer-dependent ferromagnetism in a van der Waals crystal down to the monolayer limit. *Nature* **2017,** *546*, 270-273.
(5) Bernasconi, M.; Marra, G. L.; Benedek, G.; Miglio, L.; Balkanski, M.; Scagliotti, M.; Julien, C.; Jouanne, M. Lattice dynamics of layered MPX, (M =Mn, Fe,Ni, Zn; X=S,Se) compounds. *Phys. Rev. B* **1988,** *38*, 12089-12099.
(6) Joy, P. A.; Vasudevan, S. Magnetism in the layered transition-metal thiophosphates MPS$_3$ (M =Mn, Fe, and Ni). *Phys. Rev. B* **1992,** *46*, 5425-5433.
(7) Lee, J. U.; Lee, S.; Ryoo, J. H.; Kang, S.; Kim, T. Y.; Kim, P.; Park, C. H.; Park, J. G.; Cheong, H. Ising-type magnetic ordering in atomically thin FePS$_3$. *Nano Lett.* **2016,** *16*, 7433-7438.
(8) Wettling, W. Magneto-optics of ferrites. *J. Magn. Magn. Mater.* **1976,** *3*, 147-160.
(9) Kuch, W.; Schäfer, R.; Fischer, P.; Hillebrecht, F. U. *Magnetic Microscopy of Layered Structures*; Springer, 2015.
(10) Wu, Y. J.; Shen, C.; Tan, Q. H.; Shi, J.; Liu, X. F.; Wu, Z. H.; Zhang, J.; Tan, P. H.; Zheng, H. Z. Valley Zeeman splitting of monolayer MoS$_2$ probed by low-field magnetic circular dichroism spectroscopy at room temperature. *Appl. Phys. Lett.* **2018,** *112*, 153105.
(11) Mason, W. R. Theoretical Framework: Definition of MCD Terms. In *A Practical Guide to Magnetic Circular Dichroism Spectroscopy*; John Wiley & Sons, Inc., 2007; pp 14-35.
(12) Sugano, S.; Kojima, N. *Magneto-Optics*; Springer Science & Business Media, 2013.